\documentclass[aps,twocolumn,showpacs,amsmath]{revtex4}
\usepackage{graphicx}
\usepackage{amsmath}

\begin{document}

\title{Universal characterizing topological insulator and topological
semi-metal with Wannier functions}

\author{Ye Xiong} 
\affiliation{Department of Physics and Institute of Theoretical Physics
  , Nanjing Normal University, Nanjing 210023,
P. R. China}
\email{xiongye@njnu.edu.cn}
\author{Peiqing Tong}
\affiliation{Department of Physics and Institute of Theoretical Physics
  , Nanjing Normal University, Nanjing 210023,
P. R. China}
\affiliation{Jiangsu Key Laboratory for Numerical Simulation of Large
  Scale Complex Systems, Nanjing Normal University, Nanjing 210023,
P. R. China}
\affiliation{Kavli Institute for Theoretical Physics China, CAS,
Beijing 100190, China} 
\email{pqtong@njnu.edu.cn}

\begin{abstract}
The nontrivial evolution of Wannier functions (WF) for the occupied
bands is a good starting point to understand topological insulator. By
modifying the definition of WFs from the eigenstates of the projected
position operator to those of the projected modular position operator,
we are able to extend the usage of WFs to Weyl metal where the WFs in
the old definition fails because of the lack of band gap at the Fermi
energy. This extension helps us to universally understand topological
insulator and topological semi-metal in a same framework. Another
advantage of using the modular position operators in the definition is
that the higher dimensional WFs for the occupied bands can be easily
obtained. We show one of their applications by schematically explaining
why the winding numbers $\nu_{3D}=\nu_{2D}$ for the 3D topological
insulators of DIII class presented in Phys. Rev. Lett. 114,
016801(2015).
\end{abstract}

\maketitle

\section{Introduction}

The discovery of quantum Hall effect \cite{PhysRevLett.45.494}, for the
first time, sheds light on the topological properties of electronic
bands hidden behind the 2-dimensional (2D) Fermi gas in magnetic field.
Thouless {\it et al.} (referred as TKNN) calculated the Hall conductance
by Kubo formula \cite{PhysRevLett.49.405} and expressed it as an
integral in the whole first Brillouin zone (FBZ). Then Simon {\it et
al.} recognized the integral as the first Chern class of the U(1)
principal fiber bundle on the torus of the FBZ \cite{PhysRevLett.51.51,
PhysRevLett.51.2167, Kohmoto}. Besides this route, Laughlin raised
another way to understand quantum Hall effect by a gauge invariant
argument (referred as the Laughlin argument on a cylinder)
\cite{PhysRevB.23.5632}. These two ways are equivalent but the latter
one is more profound in explaining the robustness of the Hall
conductance against local disorders because it is based on the global
gauge arguments of the 2D electron system.

For topological insulators (TI), e. g., 2D spin Hall system
\cite{PhysRevB.74.195312} and 3D strong or weak TI
\cite{Bernevig15122006, RevModPhys.82.3045, TIbook}, there are also two
routes, parallel to those for quantum Hall effect, to understand
their topological properties. One is the topological invariant,
likes the Hall conductance in the TKNN theorem, that can be expressed as
an integral in a portion of the FBZ \cite{ droplet, PhysRevB.78.195125,
RevModPhys.80.1083, TIbook, RevModPhys.82.3045, RevModPhys.83.1057}.
The other route, similar to the Laughlin argument, is based on the
evolution of the Wannier centers (WC) of the localized Wannier functions
(WF) (along one direction) as a function of the momentum (in the
perpendicular directions). These two routes are also equivalent. For
instance, in Ref. \cite{PhysRevB.74.195312}, the $Z_2$ topological
invariant in the 2D spin Hall system is defined by the switch partners
of the WFs, which can be schematically illustrated by the evolution of
WCs in the $x$ direction as a function of momentum in the $y$ direction,
$k_y$. In Ref.  \cite{PhysRevB.83.235401, PhysRevB.84.075119,
PhysRevB.89.115102}, the topological invariant for 2D and 3D TIs are
also associated with the evolution of the WCs as a function of $k_y$
(and $k_z$) as well. Beside the schematic illustration of the nontrivial
topological bands, WFs are also a good one-particle basis to study the
many-body wave functions of the fractional topological insulator by
mimicking the combination of these WFs as that of the Landau orbitals
playing roles in the Laughlin state in the fractional quantum Hall effect
\cite{PhysRevLett.107.126803, PhysRevB.86.085129}, or introducing pump
process to axion coupling directly.  \cite{PhysRevLett.114.096401}

Besides TI, another group of topological materials, called
topological semi-metal (TM), such as Weyl metal, has attracted great
interest for their exotic surface states and bulk behaviors in
recent years\cite{Search2007, PhysRevB.83.205101,
PhysRevLett.107.127205}.  Weyl metal does not have topological invariant
defined in the 3D FBZ because bands touch at the Weyl points so that
there is no band gap separating the occupied and the empty bands.
One usually needs to split the 3D FBZ into a series of 2D slides and the
topological invariant, like Chern numbers, can only be defined on these
2D slides that do not cross the Weyl points. It is also impossible to
directly adopt the evolution of WCs to understand TM because the
discussion of WCs has a prerequisite condition: the WCs of WFs must be
meaningful. But as discussed by Kivelso et al., the WFs became extended
as long as the band gap at the Fermi energy became zero
\cite{PhysRevB.26.4269, ThoulessWannier, PhysRevB.56.12847,
PhysRevB.74.235111}.  So the traditional WCs of WFs become meaningless
in Weyl metals.

In this paper, we adopt the modular operators, e.g. $\hat x = {\hat
x}_{mod} + {\hat N}_x L$, where $\hat x$ is the position operator in the
$x$ direction and ${\hat x}_{mod}$ is the modular position operator,
which can be expressed as $x_{mod}= x \, \text{mod} \, L$ in the
representation of $\hat x$ \cite{QP}. Here $L$ is the length of a unit
cell. We change the definition of WFs, $| \text{WF} \rangle$, from the
eigenstates of $\hat P \hat x \hat P$ to those of $\hat P {\hat x}_{mod}
\hat P$, where $\hat P$ is the project operator on the occupied
states. The WCs of these WFs measured in each unit cell are the
eigenvalues of the projected modular position operator. We find that the
standard deviation of projected modular position operator, $\langle
\text{WF} |\hat P {\hat x}^2_{mod} \hat P | \text{WF} \rangle$ is zero
for every WF $ | \text{WF} \rangle$ even when the gap between the
occupied bands and the empty bands goes to zero. As a result, the WCs of
these newly defined WFs in each unit cell are still meaningful in both
TM and TI. By this way, we can universally illustrate the
topological properties of TI and TM by evaluating the evolution of WCs.

Although the physical model may be in 2D and 3D, the WFs defined above
are 1-dimensional (1D) as they are the eigenstates of the operator $\hat
x_{mod}$ and the momenta $k_y$ and $k_z$ are only taken as parameters.
We find that such WCs of the 1D WFs are equivalent to the phase
accumulated by the Bloch states along the Wilson loop in the $k_x$
direction, which has been introduced by Yu et al. and Soluyanov et al.
individually \cite{PhysRevB.84.075119, PhysRevB.83.235401}. One may
regard the extension of the 1D WFs as a new representation of their
conclusions, but we should emphasize that beside the application of the
1D WFs to TM, from the present starting point the 2D or higher
dimensional WFs can be easily obtained. Previously, such higher
dimensional WFs are seldom used in TI due to the reasons such as $[\hat
P {\hat x} \hat P , \hat P {\hat y} \hat P] \ne 0$ that causes the
absence of 2D WFs as common eigenstates of $P {\hat x} \hat P$ and $P
{\hat y} \hat P$.  As a contrast, we will show that such obstacle is
removed in our consideration because $[\hat P {\hat x_{mod}} \hat P ,
\hat P {\hat y_{mod}} \hat P] = 0$.  Here $\hat y$ and ${\hat y}_{mod}$
are the position and modular position operators in the $y$ direction,
respectively.

This paper is organized as the following: In section \ref{sec2}, we show
that the states in each band are also the eigenstates of the modular
position operators with a certain degenerated eigenvalue $x_{mod}$. We
then show that 3D TI, as well as 3D TM, can be universally illustrated
in the same framework with ($x_{mod}$,$k_y$,$k_z$). In section
\ref{sec3}, we extend our WFs from 1D to higher dimensions. We show
their applications by illustrating in a plausible way that why the
winding numbers $\nu_{2D}=\nu_{3D}$ in two models of the DIII class. In
section \ref{sec4} we present the conclusions.  

\section{1D modular wannier functions in multi-band systems.}\label{sec2} 

Traditionally, the 1D WFs for the occupied bands are defined as the eigenstates of the
projected position operator,
\begin{equation}
  \tilde{x} = \hat P \hat x \hat P,
  \label{tildex}
\end{equation}
where $\hat x$ is the position operator in the $x$ direction and $\hat
P= \sum_{\epsilon_i <0} |\psi_i\rangle \langle \psi_i|$ is the project
operator on the occupied states in the Hilbert space. Here, 1D refers to
the direction of position operator $\hat x$, while the momenta along the
other dimensions in the 3D space of the model are taken as parameters.
Without loss of generality, we take the Fermi energy as energy zero.
These WFs and the corresponding WCs can be denoted as $|x_i \rangle$ and
$x_i$, respectively. The second moment of the WF $l^2 = \langle x_i
|(\hat x -x_i)^2| x_i \rangle \sim \frac{1}{E_g}$, where $E_g$ is the
band gap at the Fermi energy, will diverge as long as $E_g \to 0$. This
is why there is no discussion made directly from the point of view of
WFs for TMs.

Next we show that this obstacle can be removed when the modular position
operator is introduced. We will first introduce the modular variances in a
1D continuum system and then applying them to 1D discrete multi-band
models.

For a 1D infinite continuum system in the $x$ direction, one can introduce the
modular position operator ${\hat x}_{mod}$ and the modular wave-vector
operator ${\hat k}_{mod}= {\hat P}_{mod}/\hbar$ as
\begin{eqnarray}
  \hat x = {\hat N}_x L + {\hat x}_{mod} \\
  \hat k = {\hat N}_k \frac{2\pi}{L} + {\hat k}_{mod},
  \label{moda}
\end{eqnarray}
where ${\hat P}_{mod}$ is the modular momentum operator in the $x$
direction and ${\hat x}_{mod}$ (${\hat k}_{mod}$) can only take the
eigenvalues in range of $[0,L)$ ($[0,\frac{2\pi}{L})$ ) in the $x$ ($k$)
representation and ${\hat N}_x$ (${\hat N}_k$) can only take integer
values \cite{QP}. Here $L$ is an arbitrary length.

The system can be translated $L$ in the $x$ direction by applying a
translation operator $e^{i\hat k L}=e^{i {\hat k}_{mod} L}$.  Similarly,
to translate the system $\frac{2\pi}{L}$ in the momentum space, the
translation operator $e^{i \hat x \frac{2\pi}{L}} = e^{i {\hat x}_{mod}
\frac{2\pi}{L}}$ is needed. It is easy to find that $[e^{i {\hat
k}_{mod} L}, e^{i {\hat x}_{mod} \frac{2\pi}{L}}] =0$. As $e^{i {\hat
k}_{mod} L}$ ($e^{i {\hat x}_{mod} \frac{2\pi}{L}}$) is a one-to-one
mapping for ${\hat k}_{mod}$ (${\hat x}_{mod}$), we conclude that
$[{\hat x}_{mod}, {\hat k}_{mod} ] = 0$ \cite{QP}. This means that the
modular variances commute although the original variances have $[\hat x,
\hat k]=\frac{i}{2\pi}$.

Such modular variances can be naturally extended to the discrete
periodic multi-band models in solid. We take $L$ as the length of a unit
cell so that $k_{mod}$ take the value in the FBZ and $G=\frac{2\pi}{L}$
is the reciprocal vector. We denote the $i$th orbital state in the $n$th
unit cell as $|n\rangle_i$ and the $\alpha$th band state as $|k
\rangle_\alpha$, with $|k \rangle_\alpha= \frac{1}{\sqrt{N}}
\sum_{n=1}^N e^{ikn L} \sum_{j=1}^m U_{\alpha j}(k) |n \rangle_j$, where
$N$ is the number of the unit cells and $U(k)$ is a unitary matrix.
There are totally $m$ orbital states in each unit cell so that there are
$m$ bands in the FBZ.  The wave-vector operator $\hat k$ and the
position operator $\hat x$ are defined in the real space by
\begin{eqnarray}
  e^{i\hat k L} |n\rangle_i = |n-1 \rangle_i \\
  \hat x |n \rangle_i = (nL+X_i) |n \rangle_i, 
\end{eqnarray}
where $X_i \in [0,L)$. As $e^{i\hat k L}=e^{i{\hat k}_{mod} L}$ and
$e^{i\hat x G} =e^{i{\hat x}_{mod} G}$, in the
reciprocal space, by applying the above equations to $|k
\rangle_\alpha$, we have
\begin{eqnarray}
  e^{i{\hat k}_{mod} L} |k\rangle_\alpha= e^{ikL} |k\rangle_\alpha \\
  e^{i{\hat x}_{mod} G} |k \rangle_\alpha = e^{i\theta_\alpha}
  |k\rangle_\alpha.
  \label{1Dtheta}
\end{eqnarray}
In the last equation, we have used the property that in the reciprocal
space, the states $|k \rangle_\alpha$ and $|k+G \rangle_\alpha$ must
represent the same quantum state with only a possible phase difference
$\theta_\alpha \in [0,2\pi)$. Here $\theta_\alpha$ is independent of $k$
because these transformations for different $|k\rangle$ will map to
the same Wilson loop with only starting point difference. Like that in
the continuum model, $[e^{i {\hat k}_{mod} L}, e^{i {\hat x}_{mod}
G}]=0$ so that ${\hat k}_{mod}$ and ${\hat x}_{mod}$ are diagonalized
simultaneously in the reciprocal space for each band. These $N$-fold
degenerated eigenvalues of ${\hat x}_{mod}$ for each band correspond
to $N$ WFs with their WCs at $x_{mod}=\theta_\alpha/G$ in each unit
cell.

As ${\hat x}_{mod}$ is diagonal for each band, one can apply the
occupied project operator $\hat P$ from the left and the right sides
straightforwardly and the second moment of the eigenstates of the
projected modular position operator, $\langle x_i| \hat P {\hat
x}^2_{mod} \hat P| x_i \rangle - \langle x_i|\hat P {\hat x}_{mod} \hat
P |x_i \rangle^2$ is fixed at zero.  We want to emphasize that in the
above derivations, we only suppose that each band is either fully
occupied or empty but do not require a finite band gap at the Fermi
energy. 

In the case of multi-bands being occupied, it will be hard to
distinguish the band index $\alpha$ in the presence of band crossing.
This difficulty can be overridden by studying the matrix $\hat P
e^{i{\hat x}_{mod} G} \hat P$. This operator represents how much phase
is accumulated for the states $|k\rangle_\alpha$ by mapping to
$|k+G\rangle_\alpha =|k\rangle_\alpha$.  One can divide this mapping
into $N_k$ segments, $F_{n_k} = \hat P e^{i{\hat x}_{mod} G/N_k} \hat P$
with $n_k \in (1,2,\cdots, N_k)$. Then the phases ${\hat x}_{mod} G$ can
be obtained by finding the phases of the eigenstates of the product
matrix $\Pi_{n_k=1}^{N_k} F_{n_k}$. The above mapping is equivalent to
the product of the non-abelian Berry connection along the ``Wilson
loop'' raised in Ref. \cite{PhysRevB.84.075119, PhysRevB.83.235401}.
And the phases $x_{mod} G=\theta_\alpha$ are equal to the phases denoted
by $\theta_m^D$ in Ref. \cite{PhysRevB.84.075119}.

Now we apply the above results to a 3D Weyl metal with the effective
Hamiltonian 
\begin{eqnarray}
  H_{W} &= &[M_0+\beta_1(\cos(k_x)+\cos(k_y)) +\beta_2 \cos(k_z)]
  \sigma_z \nonumber \\ & &+\sin(k_x)\sigma_x +\sin(k_y) \sigma_y ,
  \label{WeylH}
\end{eqnarray}
where $\sigma_{x(yz)}$ is the Pauli matrix along $x$($yz$) direction.
Here we fix $\beta_1=0.2$ and $\beta_2=0.4$.

\begin{figure}[ht]
  \centering
  \includegraphics[width=0.45\textwidth]{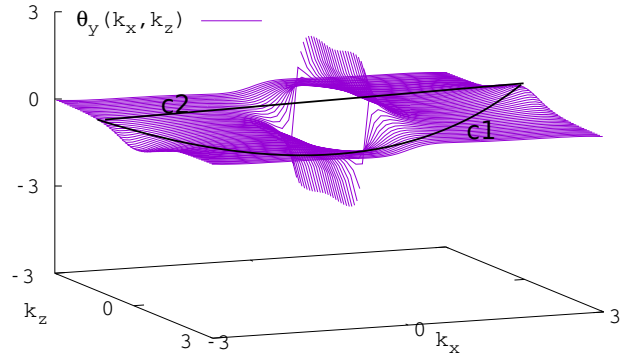}
  \caption{$\theta_y$ as a function of $k_x$ and $k_z$. $\theta_y$
  acquires a $2\pi$ ($0$) variation when $(k_x, k_z)$ evolves
along $c_2$ ($c_1$). Here $N_k=20$.}
  \label{fig2_1}
\end{figure}

The model has two Weyl points at $(k_x,k_y,k_z)=(0,0,\pm1.29)$
when $M_0=-0.5$.  Like the applications in 3D TI
\cite{PhysRevB.89.115102}, we can take the two wave-vectors, e.g. $k_x$
and $k_z$, as parameters and find the eigenvalues of $\hat P {\hat
y}_{mod} \hat P = \theta_y/G$ along the rest $y$ direction. In Fig.
\ref{fig2_1}, we plot the function $\theta_y(k_x,k_z)$ in the range
$[-\pi,\pi]$. $\theta_y$ is $2\pi$ folded along the line connecting the
two Weyl points (projected on the $k_x-k_z$ plane) which is the
projection of the Dirac string on the $k_x-k_z$ plane. Imagine that we
are varying the wave-vectors along the typical paths $C_1$ and $C_2$
shown in Fig. \ref{fig2_1}.
The WCs of WFs will come back to their original positions when the
varying is along $C_1$. But the WCs change one unit cell when the
varying is along $C_2$.  By applying Laughlin's cylinder argument, we
can conclude immediately that there should be boundary states that
cross the zero energy on the $xz$ surface to compensate the movements
of WFs to make the situation totally unchanged when the wave-vector
comes back along $C_2$ on the 2D FBZ torus. But such zero energy surface
state does not exist when the varying is along $C_1$. So these zero
energy boundary states form the unclosed Fermi arc on the $x-z$ surface
of the 3D Weyl metal. 

We find a similar plot for $\theta_x(k_y,k_z)$, which indicates that
Fermi arc can also appear at the $y-z$ surface. But we find a trivial
$\theta_z(k_x,k_y)$, keeping at $0$ for any $k_x$ and $k_y$, which
indicates that there is no Fermi arc at the $xy$ surface of the Weyl
metal.

\begin{figure}[ht]
  \centering
  \includegraphics[width=0.45\textwidth]{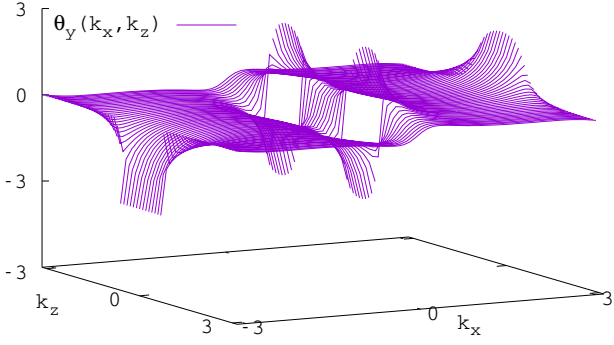}
  \caption{$\theta_y$ as a function of $k_x$ and $k_z$ for the model
    with totally $6$ Weyl points.}
  \label{fig2_2}
\end{figure}

The model can have $6$ Weyl points at $(0,\pi,\pm 0.73)$, $(\pi,0,\pm
0.73)$ and $(0,0,\pm 1.82)$ when $M_0=-0.3$. We plot the function
$\theta_y(k_x,k_z)$ in Fig. \ref{fig2_2}. As that in Fig. \ref{fig2_1},
the plot illustrates the topological properties of the Weyl metal
explicitly. The evolution of $\theta_x(k_y,k_z)$ is also similar to
$\theta_y(k_x,k_z)$ and $\theta_z(k_x,k_y)$ is trivial as well.

By the two illustrative models, we show the power of the WFs defined by
modular position operator. For the first time, we can understand TM by
considering the 3D FBZ as a whole, otherwise one usually needs to employ a
series of 2D planes or 2D cylinders in the FBZ to discuss topological
properties of TM. 

\section{Extended wannier functions to higher dimensions}\label{sec3}

In this section, we show that with the help of the modular position
operators, we can easily find WFs in higher dimensions. We will show
the application of 2D WFs in two 3D illustrative lattice models in the
DIII class.

By applying Eq. \ref{1Dtheta}, we have 
\begin{eqnarray}
  e^{i{\hat x}_{mod}G_x}|k_x,k_y \rangle_\alpha =e^{i\theta^x_\alpha(k_y)}
  |k_x,k_y \rangle_\alpha \\
  e^{i{\hat y}_{mod}G_y}|k_x,k_y \rangle_\alpha =e^{i\theta^y_\alpha(k_x)}
  |k_x,k_y \rangle_\alpha,
  \label{2Dmodular}
\end{eqnarray}
where $|k_x,k_y\rangle_\alpha$ denotes the eigenstate in the $\alpha$th
band with momenta $k_x$ and $k_y$ in the 2D FBZ. Here,
$\theta^{x(y)}_\alpha(k_{y(x)})$ is a function of $k_{y(x)}$ so that
$x_{mod}$ and $y_{mod}$ (the eigenvalues of the modular operators) are
not uniform in each unit cell. This is the key difference between WFs in
higher dimensions and those in 1D.

As the modular position operators, ${\hat x}_{mod}$ and ${\hat
y}_{mod}$, are diagonal for each band, we have $[\hat P{\hat x}_{mod}
\hat P, \hat P {\hat y}_{mod} \hat P]=0$. But for the position
operators, one has  $[\hat P{\hat x} \hat P, \hat P {\hat y} \hat P] \ne
0$ \cite{PhysRevB.56.12847}. This is why after defining the WFs by the
modular position operators, we can extend WFs to higher dimensions
easily.

We next show how to find the WCs of the WFs in 2D. The method is
straightforwardly to be applied in further higher dimensions. We take a
square 2D FBZ as an instance. We discretize the FBZ into $N^2$ elements by
$N$ rows and $N$ columns and denote the state in each cell as
$|n_x,n_y\rangle_\alpha = |-\frac{G_x}{2} +n_x \delta k_x,
-\frac{G_y}{2} +n_y \delta k_y \rangle_\alpha$, where $G_{x(y)}$ is the
reciprocal vector and $\delta k_{x(y)}=G_{x(y)}/N$. We calculate the
operators 
\begin{eqnarray}
  XY^{+}=\hat P e^{i (\hat x \delta k_x + \hat y \delta k_y)} \hat P \\
  XY^{-}=\hat P e^{i (\hat x \delta k_x - \hat y \delta k_y)} \hat P,  
\end{eqnarray}
in the discrete FBZ. Most elements of $XY^{\pm}$ are zero except
$XY^{+}_{n_x,n_y;n_x+1,n_y+1}$ for $XY^{+}$ and $XY^{-}_{
n_x,n_y;n_x+1,n_y-1}$ for $XY^-$. Furthermore, $XY^\pm$ can be block
diagonalized into $N$ independent $N\times N$ matrix, with each matrix
mimics the $\hat X_P(k_y)$ matrix in Eq. 6 in Ref.
\cite{PhysRevB.84.075119},
\begin{equation}
  XY^{\pm }(n) = \begin{pmatrix} 0 & XY^{\pm}_{1,n;2,n\pm1} & 0 &
    \cdots \\
    0 & 0 & XY^{\pm}_{2,n\pm 1;3,n\pm 2} & \cdots \\
    \hdotsfor{4} \\
    XY^{\pm}_{N,n\pm N;1,n} & 0 & 0 & \cdots \end{pmatrix}. 
  \label{XYmatrix}
\end{equation}
The phases of the eigenvalues of the product matrix
$XY^{\pm}_{1,n;2,n\pm1} XY^{\pm}_{2,n\pm 1;3,n\pm 2} \cdots
XY^{\pm}_{N,n\pm N;1,n}$ give the $N$-fold degenerated values
$x_{mod}G_x\pm y_{mod} G_y$ (mod $2\pi$). Like the 1D case, these
$N$-fold degenerated modular positions correspond to WCs localized at
different unit cells. But as there are $N$ independent $XY^{\pm}(n)$, we
will have totally $N$ kinds of $N$-fold degenerated values
$x_{mod}G_x\pm y_{mod} G_y$ (mod $2\pi$). So we can conclude that the 2D
WCs defined by these modular position operators are not uniform in each
unit cell in the 2D lattice. This is consistent with the previous
argument that the 2D WFs with uniform WCs do not exist in 2D topological
nontrivial lattice \cite{ThoulessWannier}. 

Now we apply the above method to find 2D WFs for two 3D
lattice models. We will take the momentum, e.g. $k_x$, as a parameter
and study how the 2D WCs evolve as $k_x$ is varying $2\pi$ in the FBZ. Such
illustration provides more information than the evolution of 1D
WCs as varying, e.g. both $k_x$ and $k_y$.

We start from a 3D illustrative model in the DIII class,
\begin{equation}
  H(\vec k)= \begin{pmatrix} \epsilon(\vec k) I & \Theta(\vec k) \\
    \Theta^\dagger(\vec k) & \epsilon(\vec k) I \end{pmatrix},
\end{equation}
with $\Theta(\vec k)=i (\vec{d}\cdot \vec{\sigma}) \sigma_y $, where the
unit matrix $I$ and Pauli matrix $\vec{\sigma} =
(\sigma_x,\sigma_y,\sigma_z)$ are all $2\times 2$
matrix \cite{PhysRevLett.114.016801}. 

As in Ref. \cite{PhysRevLett.114.016801}, we denote the Hamiltonian as
$H_1$ when $\epsilon(\vec k)= \cos(k_x-k_z)+\cos(k_x)+2\cos(k_y-k_z)
+\cos(k_y)-\mu$ and $(d_x, d_y, d_z)=\Delta ([\sin(k_x -k_z)+\sin(k_x)
+2 \sin(k_y-k_z)-\sin(k_y)], 2\sin(k_y), 2\sin(k_x+k_y))$. The winding
number $\nu_{3D}$ for this model can take nontrivial value $\pm 1$
\cite{PhysRevLett.114.016801}. Here we take $\Delta=5$ and $\mu=2.5$ so
that $\nu_{3D}=1$ in our calculation.  Finch et. al argued in Ref.
\cite{PhysRevLett.114.016801} that when a breaking symmetry term is
added to the Hamiltonian on the surface of this TI, e.g. on the $xy$
surface, to open up a gap for the surface states, one is able to obtain
the winding number for the gapped surface states, $\nu_{2D}$, and found
that $\nu_{3D}=\nu_{2D}$. 

\begin{figure}[ht]
  \centering
  \includegraphics[width=0.45\textwidth]{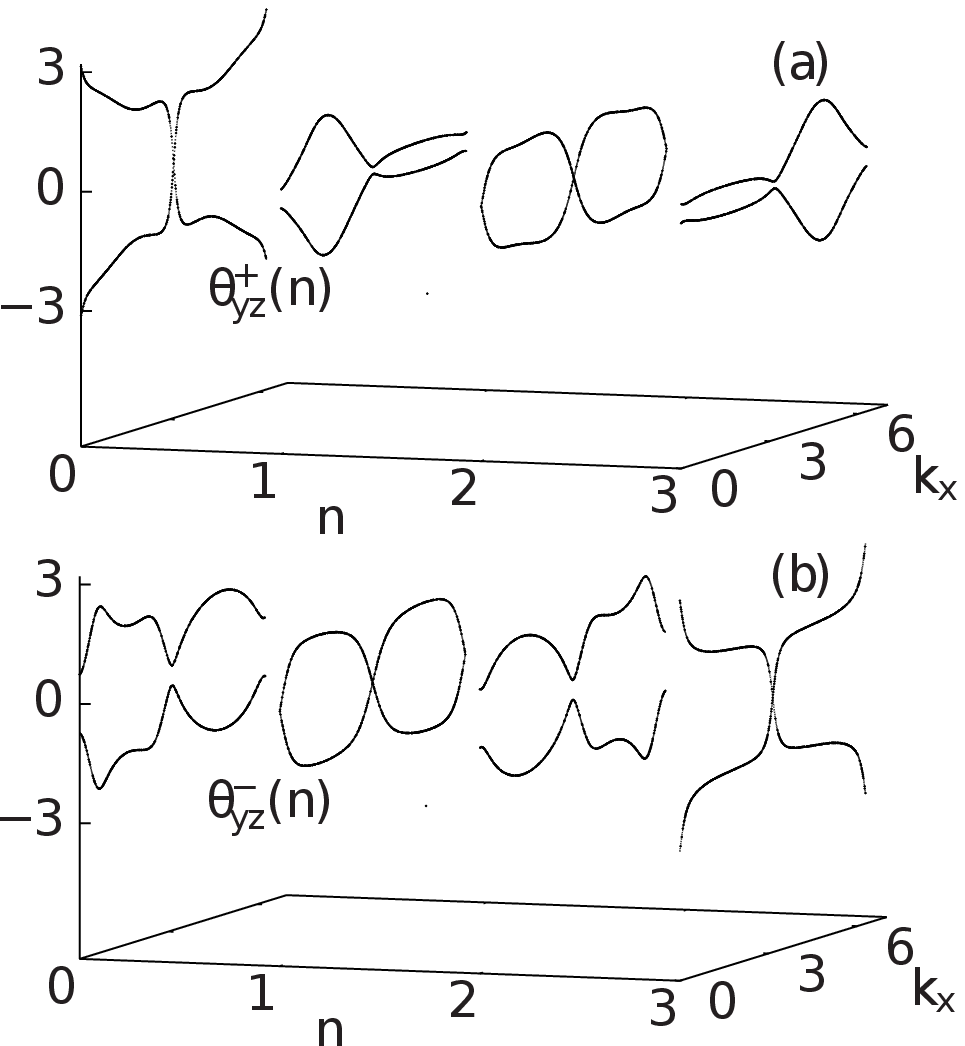}
  \includegraphics[width=0.45\textwidth]{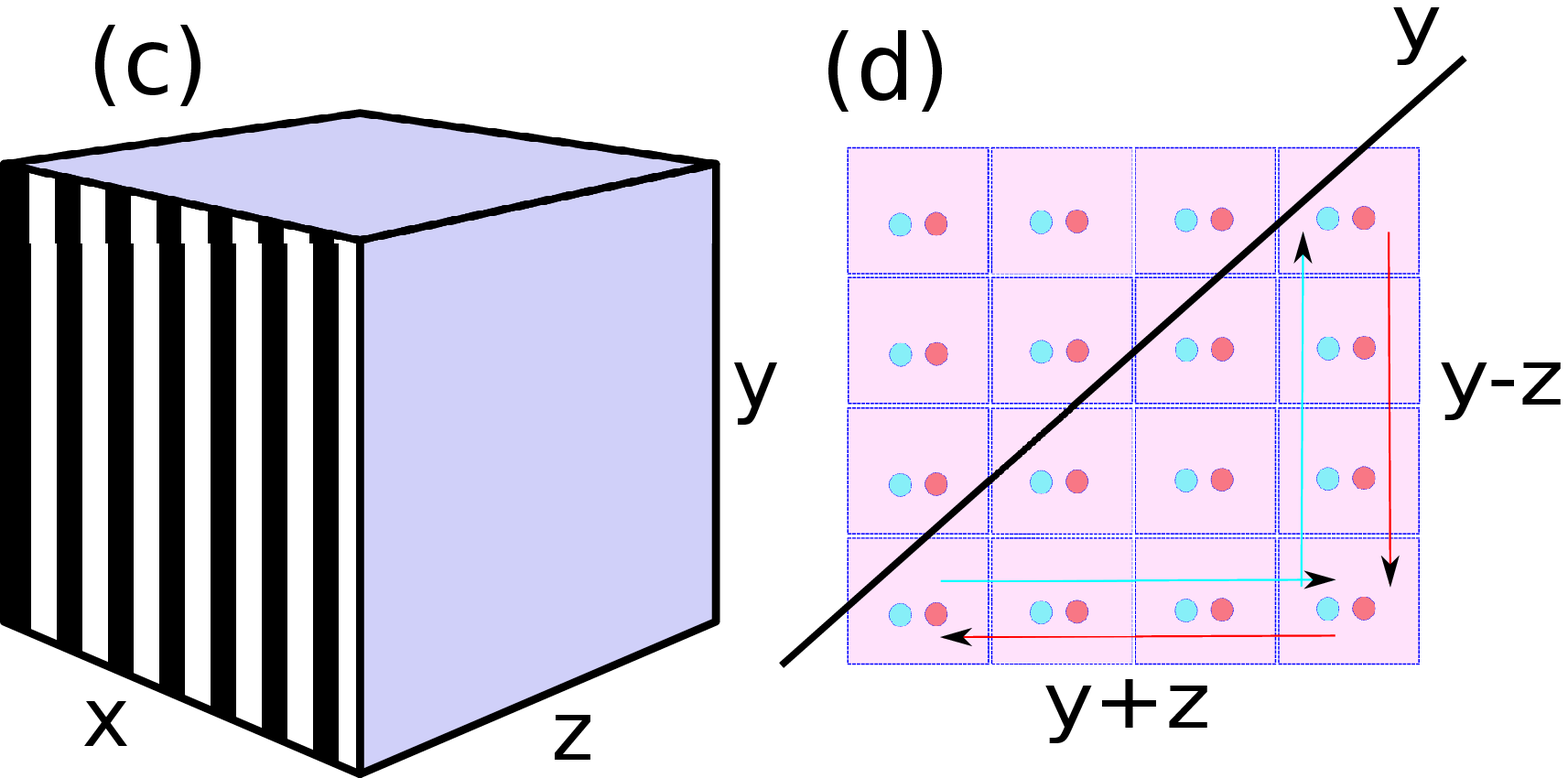}
  \caption{$\theta^{\pm}_{yz}=G(y_{mod} \pm z_{mod})$ as functions of
$k_x$ in (a) and (b). The 2D FBZ in $k_y$ and $k_z$ plane is divided in
to $4\times 4$ blocks. $n=0,1,2,3$ are the $4$ kinds of
$\theta^{\pm}_{yz}$. Here the lattice constant $a$ is taking as the unit. 
(c) shows the 3D topological insulator with
breaking symmetry term added on the $xy$ surface. We show how the WCs
of WFs evolve with $k_x$ in the 2D real space in (d), with red and green
dots representing the two WCs in each unit cell. The arrows indicate the
variations of WCs by changing $k_x$ from $0$ to $2\pi/a$. Most of them undergo a
trivial evolution except those along a column and a row.}
  \label{fig3_1}
\end{figure}

In Fig. \ref{fig3_1} (a) and (b), we illustrate how
$\theta^{\pm}_{yz}(n) = G(y_{mod}\pm z_{mod})$ evolve with $k_x$, where
$G=G_x=G_y=\frac{2\pi}{a}$ is the reciprocal vector and
$y_{mod}(z_{mod})$ is the eigenvalues of the modular position operators
$\hat y_{mod} (\hat z_{mod})$. Here, for the sake of clarity, we divide
the whole FBZ into $4\times 4$ blocks so that there are $4$ kinds of
$\theta^{\pm}_{yz}(n)$, which are identified by the index $n=0,1,2,3$ in
the figure. Dividing the FBZ into smaller parts will not alter the key
features of the figures. As the figure shows, most of the modular
positions experience a trivial evolution with $k_x$. But there is one
modular position (one for $y_{mod}+z_{mod}$ and one for
$y_{mod}-z_{mod}$ respectively) that undergoes a change of unit cell $a$
as varying a reciprocal vector for $k_x$. 

Such behavior of evolution reflects the bulk topological properties of the 3D
model which can not be modified by adding a symmetry breaking term on the $xy$ surface. 
We explicitly illustrate how these WCs evolves with
$k_x$ in the 2D real plane $yz$ in Fig. \ref{fig3_1} (d). Points in each
unit cell are representing the WCs of the WFs. The exact order of
columns or rows are not important because a gauge transformation can
exchange them. This figure shows that there is one  WF that being pumped
one half lattice constant from left to right (and another WF that is
pumped one half lattice constant inversely) along $y$ axis
as varying $k_x$ circularly from $0$ to $2\pi/a$.  To compensate such
evolution for the gauge invariant, there should be topological protected
edge states at the edge of either gapped or gapless $xy$ surface.
Because of the bulk-boundary correspondence for TI, each $xy$ surface
should contribute $1/2$ winding, which takes the responsibility of the
half lattice movement of the WFs during the $k_x$ pump. As a result,
$\nu_{2D}$, which counts the total winding of the two $xy$ surfaces, is
totally $1$.  This is what $\nu_{3D}=\nu_{2D}$ means for this model.

\begin{figure}[ht]
  \centering
  \includegraphics[width=0.45\textwidth]{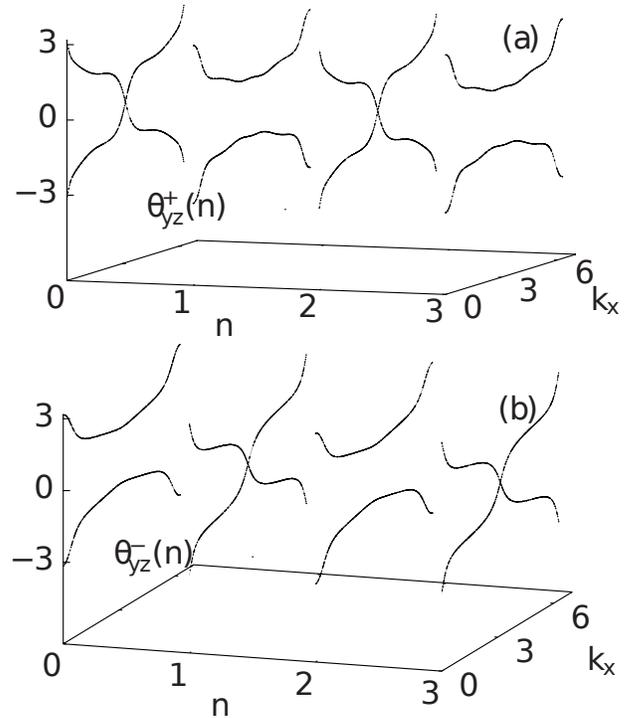}
  \caption{$\theta^{\pm}_{yz}=G(y_{mod}+z_{mod})$ as functions of
$k_x$ in (a) and (b). The 2D FBZ in $k_y$ and $k_z$ plane is divided in
to $4\times 4$ blocks. $n=0,1,2,3$ are the $4$ kinds of
$\theta^{\pm}_{yz}$.}
  \label{fig3_2}
\end{figure}

We then do a similar discussion on another model $H_2$ with $\epsilon(\vec
k)=\cos(k_x)+\cos(k_x-k_z)-\mu$ and $(d_x,d_y,d_z)=
\Delta(\sin(k_x)+\sin(k_x-k_z), 2\sin(k_y), 2\sin(k_x-k_z))$. We take
$\Delta=2$ and $\mu=1$ in the calculation. The model is in the
topological phase with $\nu_{3D}=2$.

In Fig. \ref{fig3_2}, we show how $G(y_{mod}+z_{mod})$ and
$G(y_{mod}-z_{mod})$ evolve with $k_x$. There are two modular positions
experience nontrivial evolution along $y+z$ or $y-z$ direction. This
makes us infer that there are two WCs being pumped half a unit cell from left to
right and another two WCs vice versa. Applying the similar discussion as
we have done for the previous model we can conclude that $\nu_{2D}=2$.
This also matches the equation $\nu_{3D}=\nu_{2D}$ for this model.

We have also applied similar calculations for $H_3$ and $H_4$ in the
topological phase with $\nu_{3D}=3$ and $\nu_{3D}=4$
\cite{PhysRevLett.114.016801}. But we find that the evolution of 2D WCs
are not simply extended as that from $H_1$ to $H_2$. Typically for
$H_4$, we find the evolution of 2D WCs are {\it all trivial}. One
possible explanation is that the evolution of 2D WFs can not universally
represent all TIs of the DIII class. But we propose that it is because
another factor which has not been discussed in Ref.
\cite{PhysRevLett.114.016801}, the disorder effect. Our discussion is
based on a gauge argument. So the conclusion naturally possesses the
effect of disorder even though the disorder term is not explicitly
engaged. For $H_4$ with $\nu_{3D}=4$, there is two pair of chiral edge
states at the boundary of the 2D $xy$ surface. Although the symmetry may
still prevent the intra-pairing back scattering, it can not prevent the
inter-pairing back scattering in the presence of disorder. So the
disorder can smear out the edge states near the zero energy and open up
an effective gap. This is why we find the trivial 2D WCs evolution for
$H_4$ and it will need further numerically investigation to be proved.

\section{Conclusions}\label{sec4}

We modify the definition of WFs for the occupied bands by introducing the
modular position operators. One can calculate these 1D WFs along the
Wilson loop along one direction of the FBZ. The topological nontrivial
band structure can be illustrated by the nontrivial evolution of WCs as
varying the momentum in the other directions. We show that such
illustration, usually only applied to TI, can also be applied to Weyl
metal in which there is no band gap at the Fermi surface. Our
illustration well explains the nontrivial band structure of Weyl metal
as well as the Fermi arc for the surface states. We also extend the 1D
WFs to higher dimensions. We show its application on explaining
$\nu_{3D}=\nu_{2D}$ in the two models belonging to the DIII class.

{\it
Acknowledgments.---} The work was supported by the State Key Program for
Basic Research of China (Grant Nos. 2009CB929504, 2009CB929501),
National Foundation of Natural Science in China Grant Nos. 10704040,
11175087.

\bibliographystyle{apsrev}
\bibliography{main}

\end{document}